\documentclass[reprint,amsmath,amssymb,prapplied,aps,superscriptaddress,nofootinbib]{revtex4-2}

\usepackage{lipsum}
\usepackage{graphicx}
\usepackage{dcolumn}
\usepackage{bm}
\usepackage[separate-uncertainty=true]{siunitx}
\usepackage[dvipsnames]{xcolor}
\usepackage{hyperref}
\hypersetup{colorlinks,allcolors=MidnightBlue}
\usepackage{url}
\usepackage{float}
\usepackage{siunitx}
\DeclareSIUnit{\belmilliwatt}{Bm}
\DeclareSIUnit{\dBm}{\deci\belmilliwatt}
\usepackage{bm}
\usepackage{multirow}

\let\oldpm\pm
\renewcommand{\pm}{\mathrel{\oldpm}}

\begin{document}

\title{Accurate gap control of 3D-integrated superconducting quantum circuits using asymmetric indium architecture and multi-angle flip-chip bonding}

\author{L.~Richard}
\author{N.~Bruckmoser}
\author{J.~Feigl}
\author{D.~Bunch}
\author{A.~Bhattacharjee}
\affiliation{Technical University of Munich, TUM School of Natural Sciences, Department of Physics, 85748 Garching, Germany}
\affiliation{Walther-Meißner-Institut, Bayerische Akademie der Wissenschaften, 85748 Garching, Germany}
\author{L.~Koch} 
\author{I.~Tsitsilin}
\affiliation{Technical University of Munich, TUM School of Natural Sciences, Department of Physics, 85748 Garching, Germany}
\affiliation{Peak Quantum GmbH, 85748 Garching, Germany}
\author{V.~P.~Bader}
\affiliation{Walther-Meißner-Institut, Bayerische Akademie der Wissenschaften, 85748 Garching, Germany}
\author{H.~Hu}
\author{L.~S\"odergren}
\affiliation{Technical University of Munich, TUM School of Natural Sciences, Department of Physics, 85748 Garching, Germany}
\affiliation{Walther-Meißner-Institut, Bayerische Akademie der Wissenschaften, 85748 Garching, Germany}
\author{S.~Filipp}
\affiliation{Technical University of Munich, TUM School of Natural Sciences, Department of Physics, 85748 Garching, Germany}
\affiliation{Walther-Meißner-Institut, Bayerische Akademie der Wissenschaften, 85748 Garching, Germany}
\affiliation{Munich Center for Quantum Science and Technology (MCQST), Schellingstra\ss{}e 4, 80799 M\"unchen, Germany}
\affiliation{Munich Quantum Valley gGmbH, 85748 Garching, Germany}

\date{\today}

\begin{abstract}
 
Flip-chip 3D integration is a key enabling technology for overcoming the routing and layout constraints of planar superconducting quantum processors, but it requires accurate and uniform control of the inter-chip spacing while preserving qubit coherence. In conventional processes, thick indium bumps are deposited symmetrically on both chips, while dedicated spacers are introduced to limit the residual tilt. Here, we present a flip-chip bonding process based on an asymmetric indium architecture, combining $20~\si{\micro\meter}$ thick indium bumps on the control chip with $200~\si{\nano\meter}$ thin indium pads on the qubit chip. Using electrical and microwave measurements, we demonstrate low-resistance and low-loss indium interconnects, and show that transmon qubits integrated with this process reach relaxation and coherence times of $T_1=(159\pm14)~\si{\micro\second}$ and $T_{\mathrm{2,echo}} = (294\pm42)~\si{\micro\second}$, respectively, comparable to planar devices. Combined with a multi-angle compression procedure that suppresses systematic tilt, this approach reduces the average post-bonding tilt to $(20\pm12)~\si{\micro\radian}$. These results establish an asymmetric indium architecture combined with multi-angle compression as a robust, coherence-preserving route toward 3D-integrated superconducting quantum processors.

\end{abstract}
 
\maketitle

\section{\label{sec1}Introduction}
 
For quantum computers to address practically relevant problems, they must scale beyond the size of current
devices~\cite{Fowler2012SurfaceCodes, Devoret2013SuperconductingCircuitsOutlook, Gidney2021RSA2048Factorization, GoogleQuantumAI2025SurfaceCodeThreshold}. In superconducting quantum processors, flip-chip 3D integration has emerged as an important approach for scaling circuits beyond planar architectures. By distributing qubit and control structures across separate chips, flip-chip devices can increase wiring density, reduce layout complexity and enable more flexible three-dimensional circuit architectures~\cite{Field2024MultichipTunableCoupler, Kosen2024SignalCrosstalk, Norris2026MultiModuleProcessor, Wang2026HighFidelityTwoQubitGates}. 
 
The main requirement to realize this technology is the reliable fabrication of superconducting chip-to-chip interconnects that provide both mechanical and galvanic contact. The material of choice for these interconnects is indium, valued for its ductility and superconductivity below $T_c\simeq3.4~\si{\kelvin}$~\cite{Rosenberg2017_3DIntegrated, Foxen2018QubitInterconnects, Kosen2022BuildingBlocks, Norris2024ImprovedParameterTargeting, Yu2024IndiumElectroplatedFlipChip, Yusupov2025FlipChipCouplingTypes, Wang2026HighFidelityTwoQubitGates}. In conventional approaches, indium is deposited on both control and qubit chips with thicknesses of several micrometers. This geometry promotes robust bonding, as opposing indium structures plastically deform during compression and fracture native oxide layers, enabling proper galvanic connection~\cite{Rosenberg2017_3DIntegrated, Kosen2022BuildingBlocks, Norris2024ImprovedParameterTargeting, Yusupov2025FlipChipCouplingTypes}. Thick indium layers, however, introduce additional fabrication constraints on the qubit chip that can affect device performance. Notably, thick lift-off masks require high-viscosity resists and extensive wafer-scale processing, increasing the risk of residual polymer contamination~\cite{Bahena2022BathLifePhotoresist, Quintana2014MicrofabricationDecoherence, Alghadeer2025NanostructuralImperfections, Weeden2026StronglyCoupledDefects}. Alternative indium interconnect geometries, such as silicon posts covered with indium or large indium microspheres, have been explored to overcome these limitations. However, post-based approaches require deep etching early in the fabrication process, potentially affecting subsequent processing steps, while microsphere-based approaches rely on manual placement, limiting scalability for wafer-scale integration~\cite{Li2021NovelBumpingMethod, Somoroff2024FluxoniumFlipChip, Mayer2025_3DIntegratedQubits, Paradkar2025IndiumMicrospheres}.

A critical requirement for scalable three-dimensional superconducting quantum processors is accurate control of the inter-chip spacing after assembly. The vacuum gap between the chips influences critical parameters such as frequency, capacitance, and coupling strength. Consequently, variations in chip separation translate directly into circuit-parameter deviations~\cite{Kosen2022BuildingBlocks, Norris2024ImprovedParameterTargeting, Li2023FlipChipResonatorDesign, Li2023OptimizingResonatorFrequencyStability, Luo2026FeedlineCoupledResonators}. For large-scale processors, such deviations can compromise calibration and degrade performance. This is particularly challenging for indium-based bonding, as the high ductility of indium makes the final gap sensitive to bump geometry and bonding conditions. Rigid spacer structures, such as resist or silicon spacers, have been introduced to address this limitation by defining a minimal inter-chip separation~\cite{Niedzielski2019SiliconHardStop, Conner2021FlipChipArchitecture, Norris2024ImprovedParameterTargeting}. However, while spacers may establish a lower bound for the vacuum gap, they do not guarantee post-bonding parallelism, leaving the spacing susceptible to residual tilt.
 
We present an asymmetric indium architecture that decouples the constraints of robust flip-chip bonding and coherence preservation by using individual fabrication flows on the control and qubit chips. To this end, the control chip incorporates $20~\si{\micro\meter}$ thick indium bumps and resist-based spacers, while the qubit chip carries $200~\si{\nano\meter}$ thick indium pads that preserve adhesion and galvanic connection, thereby minimizing additional processing on the qubit chip. Additionally, a multi-angle compression procedure suppresses residual tilt and enables precise control of the post-bonding inter-chip spacing.

\section{\label{sec2}Sample fabrication}
 
\begin{figure}[t]
\includegraphics{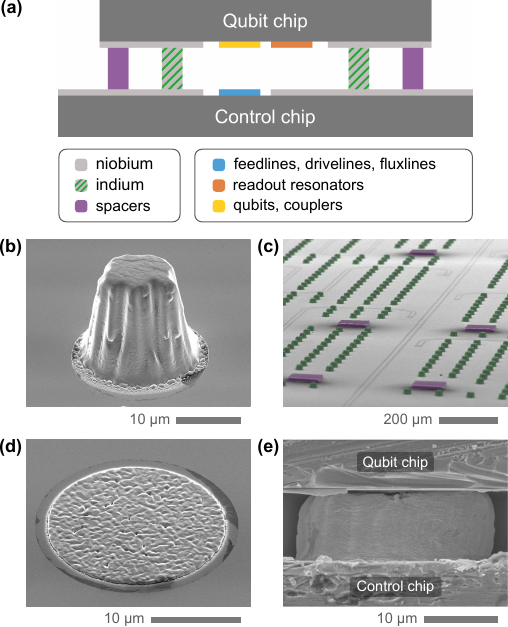}
\caption{Flip-chip module. (a) Schematic of a flip-chip structure in which a control chip and a qubit chip are connected through indium bumps and separated by polymer spacers. The control chip contains the feedlines, drivelines, and fluxlines, while the qubit chip hosts the readout resonators, qubits, and couplers. (b) SEM image of a $20~\si{\micro\meter}$ thick indium bump. (c) False-colored SEM image of a typical control-chip ground plane, with indium bumps (colored green) and spacers (colored purple). (d) SEM image of a $200~\si{\nano\meter}$ thick indium pad. (e) SEM image of an indium interconnect.}
\label{fig1}
\end{figure}

A typical flip-chip module, as illustrated in Fig.\,\ref{fig1}\,(a), consists of two chips mechanically and galvanically connected using indium bumps, and separated by rigid spacers. The control chip carries feedlines, drivelines and fluxlines, and the qubit chip hosts couplers, qubits and readout resonators. The fabrication process of each die, described in more detail in Appendix~\ref{app1}, is designed to preserve high coherence on the qubit chip by restricting resist-based spacer fabrication and thick indium evaporation to the control chip. 
 
Both dies are fabricated on separate double-side-polished silicon wafers with a diameter of $100~\si{\milli\meter}$. The silicon substrates are first cleaned in a Piranha acid solution to remove organic contaminants, followed by immersion in a buffered oxide etch (BOE) solution to remove native silicon oxide~\cite{Wisbey2010MetalSubstrateInterfaces, Martinis2012LocalizationMitigationLoss}. The wafers are subsequently transferred to a sputtering system, where a $150\,\si{\nano\meter}$ thick niobium film is deposited. For each wafer, large-area structures including coplanar waveguides and qubit capacitor pads are patterned in the niobium film using optical lithography and reactive-ion etching.
 
\begin{figure*}[t]
\includegraphics[width = \textwidth]{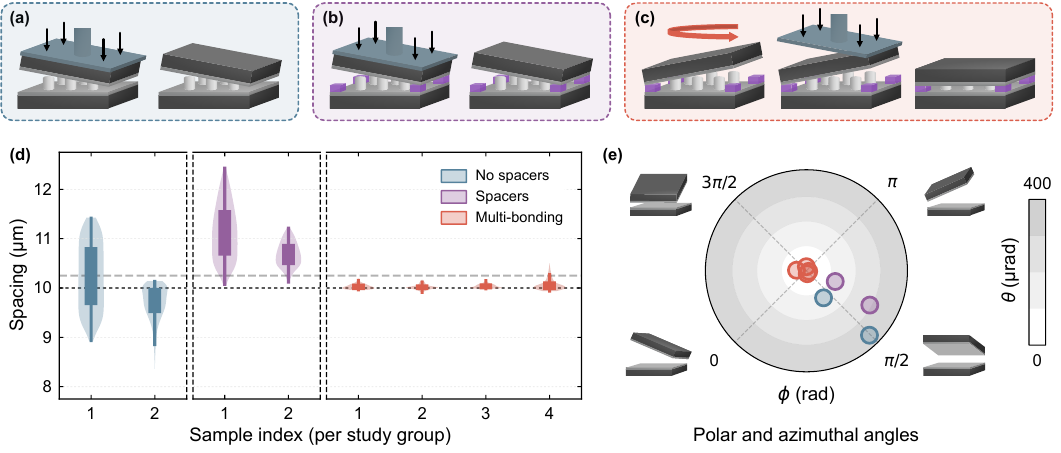}
\caption{Comparison of different bonding approaches. Group (a) Single-bonding process: control and qubit chips are bonded in a single step. Group (b) Single-bonding process including spacers: chips are bonded in a single step; spacers act as hard stops to define a minimum spacing. Group (c) Multi-angle bonding approach including spacers: after the first bond (see (b)), the sample is rotated and compressed again to compensate system-induced tilt. (d) Measurements of the inter-chip spacing for each bonding method shown as violin plots with corresponding quartiles. The gray horizontal line indicates the average spacer height, $h = 10.28\pm0.07~\si{\micro\meter}$, and the $10~\si{\micro\meter}$ spacing target is represented in black. (e) Polar projection of the tilt angles after flip-chip bonding for different study groups. The azimuthal angle $\phi$ is shown along the angular axis, while the polar angle $\theta$ is represented radially, with a gray-scale gradient indicating the magnitude of the tilt angle.}
\label{fig2}
\end{figure*}
 
For the control chip,  a layer of SU-8 3010 resist is first spin-coated and patterned by optical lithography to form spacers, with a target thickness of $10~\si{\micro\meter}$. Then, to define the indium lift-off mask, a thick negative resist is spin-coated, yielding a total thickness exceeding $22~\si{\micro\meter}$. The resist mask is patterned using optical lithography, after which the wafer is loaded into a thermal evaporator. An in-situ argon ion milling step removes surface niobium oxide to ensure low-resistance galvanic contact with the indium film. Indium is then thermally evaporated to a total thickness of approximately $20~\si{\micro\meter}$ and excess material is removed by lift-off, revealing a ground plane including $20~\si{\micro\meter}$ thick indium bumps and rigid SU-8 spacers as shown in Fig.\,\ref{fig1}\,(b) and (c). Note that all these steps must be performed at wafer scale to ensure uniform coating of the thick photoresist layers used in the process.
 
On the qubit chip, the Josephson junctions and bandages are fabricated first. Then, a resist mask for indium contact pads is defined using a thin negative resist and optical lithography. The sample is then loaded into a thermal evaporator where it undergoes in-situ argon ion milling, followed by indium evaporation. Approximately $200~\si{\nano\meter}$ of indium is deposited on the substrate, after which the excess material is removed by lift-off, leaving an indium pad, shown in Fig.\,\ref{fig1}\,(d), as a counterpart to the indium bumps on the control chip.
 
After these separate processing steps, the control and qubit chips are mounted in a flip-chip bonder (Finetech Fineplacer Femto2), one on a plate and the other on the tool's arm. After optical alignment, the chips are bonded together by applying pressure via the machine's arm, inducing mechanical compression of the indium as shown in Fig.\,\ref{fig1}\,(e).
 
\section{\label{sec3}Inter-chip spacing control}
 
To investigate the influence of spacers and bonding procedures on spacing accuracy, we compare three strategies, as illustrated in Fig.\,\ref{fig2}\,(a)\,--\,(c). Group (a) consists of samples bonded without spacers, using a single compression step. Group (b) is bonded using the same single-compression step, but incorporates SU-8 spacers with an average height of $10.28\pm0.07~\si{\micro\meter}$ across a 4-inch wafer. Group (c) includes the same spacers and is bonded using a multi-angle compression sequence. For the latter multi-angle compression sample group, a first compression step is performed, identical to the single-compression procedure. The bonded assembly is then manually rotated using cleanroom tweezers and compressed again. This sequence is repeated at nominal rotation angles of $\pi/2$, $\pi$, and $3\pi/2$ relative to the initial orientation, resulting in four compression steps in total. For each group, the initial chip orientation prior to bonding is kept identical, enabling direct comparison of the resulting tilt orientations. All bonding steps are performed at room temperature with a normal force of $16~\si{\newton}$.
 
For all three bonding strategies, spacing accuracy is characterized using dedicated test samples. The control chips measure $6\times10~\si{\milli\meter\squared}$, while the qubit chips, smaller along one dimension, measure $6\times8~\si{\milli\meter\squared}$. These dimensions are selected to produce coincident chip edges after bonding. The resulting geometry enables direct scanning electron microscopy (SEM) measurement of the inter-chip spacing, as both chip edges can be simultaneously resolved within the SEM depth of field. In addition, profilometry measurements of the qubit-chip back side after assembly are used to reconstruct the spacing over the full bonded area as SEM measurements are limited to the edges of the chip. From the reconstructed spacing map we extract the average chip spacing and its spatial standard deviation. Additionally, we extract the azimuthal angle $\phi$, which defines the out-of-plane tilt direction, and the polar tilt angle $\theta$, which quantifies the out-of-plane tilt amplitude, as detailed in Appendix~\ref{app2}.
 
The resulting spacing of the three bonding procedures is summarized in Fig.\,\ref{fig2} (d) and (e). Samples bonded without spacers (group (a)) reach the target spacing of $10~\si{\micro\meter}$ on average, with a mean spacing of $9.97~\si{\micro\meter}$, but exhibit substantial spatial variation, with a root-mean-square (rms) spatial standard deviation of $511~\si{\nano\meter}$. This variation is primarily attributed to chip tilt, as reflected by the average polar angle of $\theta_a=(248\pm118)~\si{\micro\radian}$. The results are consistent with previously reported tilt angles of samples bonded without spacers~\cite{Kosen2022BuildingBlocks, Norris2024ImprovedParameterTargeting}. For spacer-bonded samples (group (b)), the minimum spacing is larger than $10~\si{\micro\meter}$, indicating that a well-defined separation is set by the spacers and that over-compression is effectively eliminated. However, spacers alone cannot compensate for under-compression of the indium bumps caused by imperfect parallelism between the chip surfaces, which can result from system-induced tilt, residual particles, or surface imperfections. This is reflected in an rms spatial standard deviation of $418~\si{\nano\meter}$ and a polar tilt angle of $\theta_b=(209\pm84)~\si{\micro\radian}$, comparable to those of samples bonded without spacers. The azimuthal angles of the single-compression sample groups, $\phi_a=(1.46\pm0.10)~\si{\radian}$ and $\phi_b=(1.93\pm0.07)~\si{\radian}$ for samples without (group (a)) and with spacers (group (b)) respectively, indicate that the tilt direction clusters around a preferred orientation of approximately $\pi/2~\si{\radian}$. This indicates the presence of a systematic, process-induced tilt offset, which, given its high reproducibility, we attribute to the flip-chip bonder. 
 
The multi-angle bonding procedure largely eliminates this systematic contribution. By averaging the system-induced deformation over four orientations, the polar tilt is reduced by more than an order of magnitude to $\theta_c=(20\pm12)~\si{\micro\radian}$. In addition, the azimuthal angles no longer exhibit a preferred direction, resulting in a distribution centered around $\phi_c=(-0.01\pm1.71)~\si{\radian}$. Correspondingly, the spacing distribution narrows substantially, with multi-angle bonded samples exhibiting a mean spacing of $10.04~\si{\micro\meter}$ and an rms spatial standard deviation of $64~\si{\nano\meter}$. The residual offset between the final bonded spacing and the original spacer height is discussed in Appendix~\ref{app3}.
 
To assess whether the multi-angle bonding procedure remains effective for different chip dimensions, we apply the same protocol to control and qubit chips both measuring $14.3\times14.3~\si{\milli\meter\squared}$. These dimensions are representative of the chip sizes required for multi-qubit superconducting circuits with up to 20 qubits. Following the multi-angle bonding of four assemblies, the reconstructed spacing maps yield an average azimuthal and polar angle $\phi_L=(-0.28\pm1.09)~\si{\radian}$ and $\theta_L=(23\pm8)~\si{\micro\radian}$ respectively. Furthermore, the samples exhibit a mean spacing of $10.08~\si{\micro\meter}$ with an rms spatial standard deviation of $158~\si{\nano\meter}$ over the full chip area. These results demonstrate that the multi-angle procedure maintains high parallelism when extended to larger-area flip-chip assemblies.
 
\section{\label{sec4}Parameter targeting}
 
Having established that the multi-angle process provides reproducible chip spacing and low residual tilt, we next evaluate whether this accuracy is sufficient to preserve circuit-parameter targeting in superconducting qubit devices. 
 
The flip-chip architecture affects the circuit parameters most notably
through the readout-resonator frequencies and the qubit capacitance. In the transmon regime ($E_J\gg E_c$), the qubit frequency is $f_q \approx \sqrt{8E_JE_c}-E_c$~\cite{Koch2007Transmon}, where the Josephson energy $E_J$ is set by the junction and the charging energy $E_c=e^2/2C_q$ is set by the total qubit capacitance $C_q$. In a flip-chip geometry, $C_q$ includes the capacitance between the qubit pads and the opposing control-chip ground plane, which depends on the inter-chip spacing. A larger spacing reduces $C_q$ and therefore increases $E_c$ and $f_q$, while $E_J$ is
unaffected. In contrast, a larger spacing increases the effective permittivity of a CPW line and lowers the resonator frequency~\cite{Li2023FlipChipResonatorDesign, Kosen2022BuildingBlocks, Norris2024ImprovedParameterTargeting}.
 
In a $6\times10~\si{\milli\meter\squared}$ and $6\times8~\si{\milli\meter\squared}$ dies geometry, the previously measured average residual tilt without multi-angle bonding, $209~\si{\micro\radian}$, would correspond to spacing variations of up to $2~\si{\micro\meter}$ above the $10~\si{\micro\meter}$ target. Simulations of a transmon design show that a $2~\si{\micro\meter}$ variation in inter-chip spacing can result in a $2\,\%$ deviation in CPW resonator frequency and a $6\,\%$ deviation in qubit capacitance. The latter corresponds to a $6.5\,\%$ variation in charging energy $E_c$ and a $2.6\,\%$ variation in qubit frequency. In comparison, a residual tilt of approximately $20~\si{\micro\radian}$ achieved with the multi-angle process corresponds to only $0.2~\si{\micro\meter}$ spacing deviation, reducing the potential absolute deviation in circuit parameter attributed to tilt below $0.2\,\%$ for the resonator and qubit frequencies, and below $1.1\,\%$ for the charging energy. This dramatically decreases the impact of flip-chip bonding on circuit parameter targeting. Details of the simulation can be found in Appendix~\ref{app5}.
 
\begin{table}[t]
\centering
\vspace{-5pt}
\caption{Comparison of the average parameter error between simulation and measurement, before and after flip-chip assembly. Here, $\overline{\epsilon}_x$ denotes the average signed percentage deviation from the simulated target value, with $\epsilon_x=100\times(x_\mathrm{meas}-x_\mathrm{sim})/x_\mathrm{sim}$. The parameters studied are the readout resonator frequency $f_r$, the qubit frequency $f_q$ and the charging energy $E_c$.}
\vspace{10pt}
\renewcommand{\arraystretch}{1.5}
\setlength{\tabcolsep}{15pt}
\begin{tabular}{ccc}
\hline
Error & 2D & 3D (multi-angle) \\
\hline
    $\bm{\overline{\epsilon}_{\mathbf{f_r}}}$ & $-4.46 \pm 0.54$ \% & $-3.92 \pm 0.34$ \% \\
    $\bm{\overline{\epsilon}}_{\mathbf{f_q}}$ & $-2.08 \pm 1.53$ \% & $-1.20 \pm 1.51$ \% \\
    $\bm{\overline{\epsilon}_{E_c}}$ & $-1.51 \pm 0.17$ \% & $-1.18 \pm 0.59$ \% \\
\hline
\end{tabular}
 
\vspace{-5pt}
\label{tab1}
\end{table}

To study the effect of multi-angle bonding on the circuit parameters, we design transmon circuits that can be measured in both a planar and a flip-chip configuration. The geometry is chosen such that the qubit frequency lies at $4.5~\si{\giga\hertz}$ in the planar and  $3.8~\si{\giga\hertz}$ in the flip-chip configuration. This allows us to compare the deviations from the respective simulated targets in both architectures, and to isolate drifts related to tilt from other effects. The sample has a chip size of $6\times10~\si{\milli\meter\squared}$ in the planar architecture, and can be subsequently bonded to a $6\times8~\si{\milli\meter\squared}$ die for the flip-chip architecture. The device, composed of six transmon qubits, is first measured in its planar geometry, where circuit parameters including readout resonator frequencies, qubit frequencies, and charging energies are extracted. The sample is subsequently assembled using the multi-angle flip-chip bonding procedure and remeasured. Table~\ref{tab1} summarizes the average deviations across all qubits for the planar and flip-chip configurations, while the complete set of individually extracted resonator and qubit parameters is provided in Appendix~\ref{app7}. After flip-chip assembly, the measured resonator frequencies are on average $3.92\pm0.34\,\%$ lower than the simulated target values. A similar deviation is observed in the planar configuration, where the measured frequencies are $4.46\pm0.54\,\%$ lower than the simulated values. The difference between the two configurations is within their respective standard deviations and is therefore not significant. The qubit frequencies and charge energies exhibit similar behavior, with deviations at the few-percent level and changes after assembly remaining below the limit set by the simulation of a $20~\si{\micro\radian}$ tilt. Overall, the comparison demonstrates that the parameter deviations before and after multi-angle compression bonding are comparable. This indicates that the flip-chip architecture with multi-angle bonding does not lead to unwanted deviations in circuit parameters. Instead, the observed deviations are likely dominated by fabrication-related effects already present before bonding, such as variations in niobium film thickness, etch depth, or Josephson junction fabrication. 
 
\section{\label{sec5}Device characterization}
 
We next evaluate the microwave performance of the asymmetric indium flip-chip architecture. Details of the measurement setup are provided in Appendix~\ref{app6}, and three-dimensional schematics of all microwave-characterization samples are shown in Appendix~\ref{app7}. First, to compare the microwave loss associated with the control and qubit chips, we characterize $\lambda/4$ coplanar-waveguide (CPW) resonators fabricated on either the control chip or the qubit chip. The sample contains four resonators on the control chip and six resonators on the qubit chip, with resonance frequencies spanning $4$\,--\,$8~\si{\giga\hertz}$. All resonators are designed for a $50~\si{\ohm}$ characteristic impedance in the flip-chip geometry, with a $10~\si{\micro\meter}$ center trace and $9~\si{\micro\meter}$ gaps. Each resonator is capacitively coupled to a common feedline on the control chip, with a coupling quality factor of $\mathrm{Q_c} \approx 5\times10^5$. 

\begin{figure}[h!]
\includegraphics{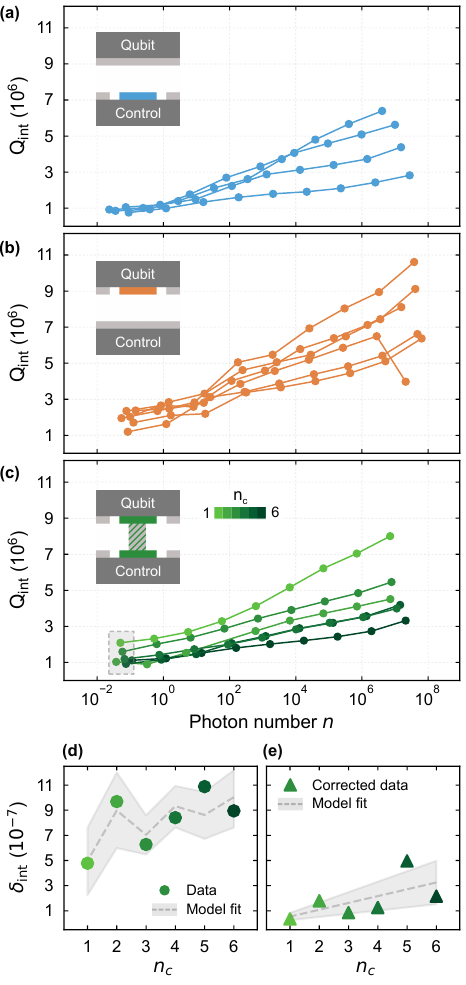}
\caption{Internal quality factors as a function of photon number for resonators in different flip-chip geometries. Error bars are included but remain below the marker size. (a) Resonators located on the control chip. (b) Resonators located on the qubit chip. (c) Resonators with their center conductor distributed between the control and qubit chips and connected with indium transitions. Here, $n_c$ denotes the number of indium transitions within the center conductor. (d) Internal loss $\delta_\mathrm{int}$ of resonators from sub-figure (c) at low photon number versus number of indium transitions $n_c$. The dashed gray line is the model of Eq.\,(\ref{eq:loss}), with the shaded area showing its 95\% confidence interval. (e) Same data after subtracting the chip-dependent term $x\,\delta_\mathrm{control}+(1-x)\,\delta_\mathrm{qubit}$. The dashed line is a linear fit through the origin, with the shaded area showing its 95\% confidence interval.}
\label{fig3}
\end{figure}

After flip-chip assembly and packaging, the sample is cooled to approximately $15~\si{\milli\kelvin}$ in a dilution refrigerator. The input power is swept from $-80~\si{\dBm}$ to $-160~\si{\dBm}$ to probe the photon-number dependence of the resonators loss. The complex transmission parameter $S_{21}$ is measured using a vector network analyzer (VNA), and the internal quality factors $\mathrm{Q_{int}}$ are extracted using a circle-fit model~\cite{McRae2020MaterialsLossMeasurements}. At high photon number, the control-chip resonators exhibit an internal quality factor of $\mathrm{Q_{int}}=(4.81\pm1.35)\times10^6$, while the qubit-chip resonators reach $\mathrm{Q_{int}}=(7.47\pm2.13)\times10^6$ on average, with the uncertainty representing the standard deviation across all resonators, as shown in Fig.\,\ref{fig3}\,(a) and (b) respectively. In the low-photon-number regime, the quality factors decrease to $\mathrm{Q_{int}}=(0.90\pm0.11)\times10^6$ and $\mathrm{Q_{int}}=(1.93\pm0.40)\times10^6$ for the control-chip and qubit-chip resonators, respectively. The reduction of $\mathrm{Q_{int}}$ at low photon number is consistent with unsaturated two-level-system (TLS) loss, as commonly observed in superconducting CPW resonators~\cite{Muller2019TwoLevelSystems}. The internal quality factors of the control-chip resonators are consistent with previously reported values for thick indium processing~\cite{Norris2024ImprovedParameterTargeting, Wang2026HighFidelityTwoQubitGates}, while the higher $\mathrm{Q_{int}}$ of the qubit-chip resonators over the full measured photon-number range indicates that the qubit-chip process yields a lower microwave-loss environment than the control-chip process. These measurement results support the conclusion that the higher internal quality factor results from the absence of additional fabrication steps on the qubit chip, such as the patterning of resist-based spacers and the deposition of a thick indium layer.
 
To enhance the routing capabilities between the control and qubit chips, indium interconnects can also be incorporated directly into the center conductor of CPW structures, forming microwave transitions between the two chips, which we refer to as indium transitions. Such a design may, however, introduce additional loss through the indium material, the indium-oxide interface, or the transition geometry. To evaluate the potential loss introduced by these indium transitions, we fabricate a second resonator sample containing six $\lambda/4$ CPW resonators in which the number of indium transitions in the center conductor ranges from one to six, numbers relevant for realistic three-dimensional routing (see layout in Appendix~\ref{app7}). At each transition, the CPW is routed from the control chip to the qubit chip, or vice versa, with the center conductor staying galvanically connected through an indium interconnect. The resonators have the same nominal CPW geometry as above, with a $10~\si{\micro\meter}$ center trace and $9~\si{\micro\meter}$ gaps, and are capacitively coupled to a common feedline with $\mathrm{Q_c} \approx 5\times10^5$. As a reference, the sample also includes four control-chip probe resonators of identical geometry but without indium transitions. In the low-photon-number regime, the control-chip resonators exhibit an average quality factor of $\mathrm{Q_{int}}=(0.97\pm0.22)\times10^6$, consistent with the control-chip resonators in Fig.\,\ref{fig3}\,(a). To quantify the contribution of the indium transitions, we restrict the following analysis to the low-photon-number data points of Fig.\,\ref{fig3}\,(c), where TLS-dominated loss is the relevant regime, and model the internal loss of each resonator as
\begin{equation}
\label{eq:loss}
\delta_\mathrm{int} = x\,\delta_\mathrm{control} + (1-x)\,\delta_\mathrm{qubit} + n_c\,\delta_\mathrm{In},
\end{equation}
where $x$ is the fraction of the CPW length located on the control chip, $n_c$ is the number of indium transitions, and $\delta_\mathrm{control}$, $\delta_\mathrm{qubit}$ and $\delta_\mathrm{In}$ are the effective loss tangents of the control chip, the qubit chip and a single indium transition, respectively. Fig.\,\ref{fig3}\,(d) compares the model evaluated at the six discrete values of $n_c$ with the measured losses. To isolate the contribution of the indium transitions, we subtract the fit-extracted term $x\,\delta_\mathrm{control}+(1-x)\,\delta_\mathrm{qubit}$ from each data point, as shown in Fig.\,\ref{fig3}\,(e)~\cite{Bruckmoser2026NiobiumAirBridges}. The least-squares fit yields $\delta_\mathrm{control}=(9.19\pm2.76)\times10^{-7}$, $\delta_\mathrm{qubit}=(3.01\pm2.19)\times10^{-7}$ and $\delta_\mathrm{In}=(0.54\pm0.42)\times10^{-7}$, with $R^2=0.68$. We attribute the moderate $R^2$ to the resonator-to-resonator scatter of the ground plane loss, even for nominally identical control-chip or qubit-chip CPWs. Additionally, owing to the small number of resonators, the uncertainties of the fit parameters are substantial. We therefore interpret the extracted values as loss upper bounds, such that we obtain $\mathrm{Q_{int}}>0.84\times10^6$ for the control chip and $\mathrm{Q_{int}}>1.92\times10^6$ for the qubit chip, consistent with the values measured in Fig.\,\ref{fig3}\,(a,b). Accounting also for the large fit uncertainty, within $1\sigma$, the indium loss tangent is bounded by $\delta_\mathrm{In}<1\times10^{-7}$ per indium transition. This is an order of magnitude smaller than the effective loss of the control-chip reference resonators, $\delta_\mathrm{ref}=(1.03\pm0.23)\times10^{-6}$. This comparison indicates that the measured loss is dominated by the control-chip microwave environment rather than by the indium transitions.

\begin{figure}[t]
\includegraphics[width=0.48\textwidth]{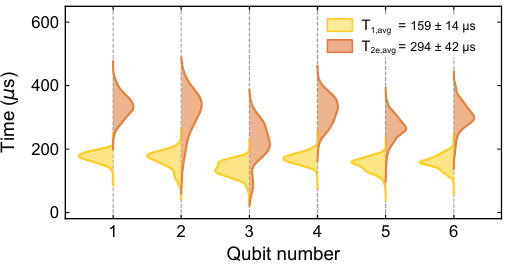}
\caption{$T_1$ and $T_{2,\mathrm{echo}}$ measurements of isolated transmon qubits. The measurements, presented in violin plots, are performed continuously over a 24-hour period.}
\label{fig4}
\end{figure}

Finally, we evaluate whether the asymmetric indium integration process preserves the coherence properties of transmon qubits using the flip-chip device introduced in Sec.\,\ref{sec4}, in which six isolated qubits exhibit an average frequency $f_q= (3.75\pm0.06)~\si{\giga\hertz}$. To capture temporal fluctuations and obtain representative coherence metrics, we repeatedly measure the relaxation time $T_1$ and echo dephasing time $T_{2,\mathrm{echo}}$ of all qubits over a 24-hour period, resulting in 100 data points per qubit and per parameter. The flip-chip device exhibits an average relaxation time of $T_{1,\mathrm{avg}}=(159\pm14)~\si{\micro\second}$ across six qubits, reaching $T_{1}$ above $200~\si{\micro\second}$, as presented in Fig.\,\ref{fig4}. These values are comparable to those of a planar transmon sample fabricated with the same process, which exhibits $T_{1,\mathrm{avg}}=(167\pm34)~\si{\micro\second}$ across six qubits. In addition, the flip-chip device achieves an average echo dephasing time of $T_{2,\mathrm{echo,avg}}=(294\pm42)~\si{\micro\second}$. The comparable coherence performance between flip-chip and planar devices demonstrates that the asymmetric indium integration process preserves transmon coherence and does not introduce significant additional loss channels within the sensitivity of the present measurements.

\section{\label{sec6}Conclusion and Outlook}
 
In conclusion, we have demonstrated a flip-chip integration process that combines an asymmetric indium geometry with a multi-angle bonding procedure to achieve accurate control of the post-bonding chip geometry. By confining the $20~\si{\micro\meter}$ thick indium bumps and $10~\si{\micro\meter}$ high SU-8 spacers to the control chip, the qubit chip requires only $200~\si{\nano\meter}$ thick indium pads and no thick-resist processing after Josephson junction fabrication. Applied in four compression steps at successive $90\si{\degree}$ orientations, the multi-angle procedure reduces the mean residual tilt from $(209\pm84)~\si{\micro\radian}$ for a single-step bond with spacers to $(20\pm12)~\si{\micro\radian}$, and lowers the rms inter-chip spacing non-uniformity from $418~\si{\nano\meter}$ to $\SI{64}{\nano\meter}$. This parallelism is preserved at processor-relevant chip dimensions of $14.3\times14.3~\si{\milli\meter\squared}$, where we measure a residual tilt of $(23\pm8)~\si{\micro\radian}$.
 
The interconnects formed by this process are both electrically robust and low-loss. Single-photon microwave characterization yields internal quality factors of $\mathrm{Q_{int}}=(1.93\pm0.40)\times10^6$ for resonators on the qubit chip and $\mathrm{Q_{int}}=(0.90\pm0.11)\times10^{6}$ for the control chip, which is exposed to more processing steps, and a loss per indium transition $\delta_\mathrm{In} < 1\times10^{-7}$. Single transmons integrated with this process reach $T_1 = (159\pm14)~\si{\micro\second}$, comparable to the $T_1 = (167\pm34)~\si{\micro\second}$ of planar reference qubits fabricated without flip-chip integration.
 
These results establish asymmetric indium bonding with multi-angle compression as a robust, coherence-preserving integration route and a scalable foundation for large-scale three-dimensional integration of superconducting quantum processors.

\begin{acknowledgments}
This work received financial support from the German Federal Ministry of Education and Research via the funding program ’Quantum technologies - from basic research to the market’ under contract number 13N15680 (GeQCoS) and under contract number 13N16188 (MUNIQC-SC), by the Deutsche Forschungsgemeinschaft (DFG, German Research Foundation) via the Germany’s Excellence Strategy EXC- 2111-390814868 ‘MCQST’ as well as by the European Union by the EU Flagship on Quantum Technology HORIZON-CL4-2022-QUANTUM-01-SGA project 101113946 OpenSuperQPlus100. The research is part of the Munich Quantum Valley, which is supported by the Bavarian state government with funds from the Hightech Agenda Bayern Plus.
\end{acknowledgments}

\appendix
 
\renewcommand{\thefigure}{A\arabic{figure}}
\setcounter{figure}{0}
 
\renewcommand{\thetable}{A\arabic{table}}
\setcounter{table}{0}
 
\renewcommand{\theequation}{A\arabic{equation}}
\setcounter{equation}{0}
 
\begin{figure*}
\includegraphics[width=\textwidth]{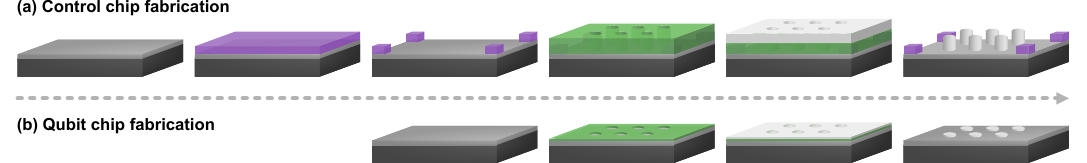}
\caption{Illustrations of flip-chip related fabrication steps for the control and qubit samples, shown in rows (a) and (b), respectively. (a) The control ground plane is first coated with SU-8 3010 resist (purple), from which spacers are patterned. The substrate is subsequently coated with AZ 15 nXT 450 cP resist and patterned optically to define a lift-off mask (green). The sample is placed in a thermal evaporator where it undergoes argon milling and indium evaporation. The resist mask and excess indium are removed by lift-off, and the spacers and indium bumps are revealed. (b) After ground plane fabrication, and if needed, Josephson junction fabrication, the qubit-chip substrate is coated with ma-N 1420 and patterned using optical lithography (green). The sample is also placed in a thermal evaporator, where argon ion milling and indium evaporation are performed. After lift-off, previously fabricated structures and indium pads are revealed.}
\label{fig_app1}
\end{figure*}
 
\section{\label{app1}Fabrication details}
 
The fabrication sequence described in Sec.~\ref{sec2} is presented here with additional process-specific information. 
 
Both the control and qubit chips are fabricated on separate double-side-polished, high-resistivity silicon wafers ($>10~\si{\kilo\ohm\centi\meter}$) with a diameter of $100~\si{\milli\meter}$. Following substrate cleaning in a 3:1 Piranha solution and a 7:1 buffered oxide etch (BOE), a $150~\si{\nano\meter}$ niobium film is deposited in an ultrahigh-vacuum sputtering system (PLASSYS MEB550 S4-I). Optical lithography is performed using a Heidelberg Instruments MLA150 maskless aligner, and the niobium layer is patterned by $\mathrm{SF_6}$ reactive-ion etching using an Oxford Instruments PlasmaPro 100 Cobra system. After ground plane etching, the control and qubit chips follow separate fabrication flows, as represented in Fig.\,\ref{fig_app1}.
 
For the control chip, the wafer undergoes an additional Piranha treatment to grow a thin oxide layer, promoting adhesion and mechanical stability of the epoxy-based resist~\cite{Lai2019SU8Adhesion}. The resist-based spacers are fabricated from SU-8 3010 photoresist, which is spin-coated, patterned by optical lithography, and developed in mr-Dev 600. To improve mechanical stability, the spacers are hard baked at $150\si{\celsius}$. The wafer is subsequently immersed in BOE to remove the oxide layer formed during the adhesion treatment and hard baked again at $150\si{\celsius}$. The thick-indium lift-off mask consists of two consecutively spin-coated layers of negative resist AZ 15 nXT 450 cP. The mask is patterned using optical lithography and developed in AZ MIF 2026. The wafer is then transferred to a dedicated thermal evaporator (PLASSYS ME450S-In), where in-situ argon ion milling is performed prior to the deposition of $20~\si{\micro\meter}$ of indium. During the evaporation, the substrate is cooled to $\num{-10}\si{\celsius}$ and indium is evaporated at a rate of around $10~\si{\nano\meter}\cdot\si{\per\second}$. Lift-off is performed first in acetone at $40\si{\celsius}$ and subsequently in Technistrip NI555 at $100\si{\celsius}$ to remove residual resist monolayers. The wafer is finally coated with a protective AZ 15 nXT 450 cP resist layer prior to dicing in an automatic wafer-dicer.
 
For the qubit chip, following ground plane patterning, the substrate is cleaned in BOE. When applicable, Josephson junctions are fabricated at this stage. To fabricate Josephson junctions, we use the Manhattan shadow evaporation process. A bilayer electron-beam resist stack consisting of a $650~\si{\nano\meter}$ thick CSAR 6200 bottom layer and a $350~\si{\nano\meter}$ thick PMMA 950K top layer is patterned by electron-beam lithography to define a cross-shaped structure required for Manhattan double-angle evaporation. In a PLASSYS MEB550 S4-I, aluminum is evaporated at a tilt angle of $\SI{35}{\degree}$ with substrate rotations of $\SI{0}{\degree}$ and $\SI{90}{\degree}$ to form the Josephson junction electrodes, with an oxidation step between the two evaporations. Following evaporation, the resist mask and excess aluminum are removed in an N-methyl-2-pyrrolidone (NMP) bath. Ex-situ bandages are subsequently fabricated using the same lithographic process, preceded by an additional argon ion milling step immediately before aluminum deposition to ensure good electrical contact. Following junction fabrication, an indium pad resist mask is defined using a $2~\si{\micro\meter}$ thick ma-N 1420 negative resist, optical lithography, and development in ma-D 533/s. The sample is then loaded into the thermal evaporator, where in-situ argon ion milling is performed before approximately $200~\si{\nano\meter}$ of indium is deposited. Again, the substrate is cooled to $\num{-10}\si{\celsius}$ and indium is evaporated at a rate of around $10~\si{\nano\meter}\cdot\si{\per\second}$. Lift-off is performed in acetone at $40\si{\celsius}$. 
 
The qubit-chip process intentionally follows a junctions-first, indium pads-second fabrication sequence. Significant deterioration of the indium pads is observed when they undergo the complete process associated with junction fabrication, notably during the lift-off. By fabricating the Josephson junctions before the indium pads, we prevent the indium from being exposed to harsh chemicals, thus eliminating the risk of degradation while remaining fully compatible with the subsequent flip-chip process. 

 
For flip-chip bonding, the control and qubit chips are mounted in a flip-chip bonder (Finetech Fineplacer Femto2). One chip is positioned face up on a polished tungsten carbide plate held in place by vacuum, while the other is placed face down on a holder, supported only at its corners. The latter is picked up by vacuum using the Femto2 arm, which is equipped with a polished stainless steel tip. Using a combination of two cameras, one oriented toward the plate and the other toward the arm, together with alignment markers patterned in the niobium layers, automatic $x$-$y$ image-registration is performed, ensuring proper alignment between the indium bumps and pads. The chips are bonded together by applying force via the arm, inducing mechanical compression of the indium. For the multi-angle bonding procedure, after the first bond, the compressed sample is picked up from the plate and rotated by hand using cleanroom tweezers, and then placed back on the plate for the next bonding step. Rotation and placement accuracy are subject to user-dependence.
 
Using infrared microscopy and dedicated cross markers before and after the additional compression steps, we investigate the effect of multi-angle compression on $x$-$y$ alignment. Within an estimated measurement accuracy of approximately $1~\si{\micro\meter}$, no measurable degradation of the in-plane displacement or relative in-plane rotational alignment is observed.

\section{\label{app2}Spacing reconstruction}
 
\begin{figure}[b]
\includegraphics{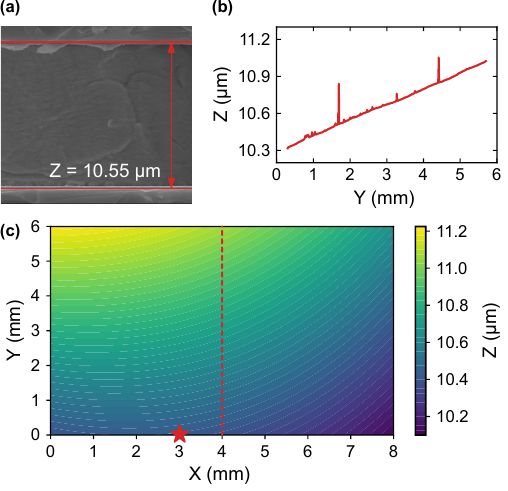}
\caption{Chip spacing reconstruction method. (a) SEM imaging along the overlapping edges of a bonded sample. The red lines highlight the silicon substrates, while the arrow represents the spacing between the silicon planes. (b) Example of a re-calibrated profilometer measurement of a top chip. (c) Spacing map extracted by quadratic fit of SEM imaging and profilometry data. The red star indicates the location of the SEM image in sub-figure (a) and the dashed red line indicates the location of the profilometry data of sub-figure (b).}
\label{fig_app2}
\end{figure}

The inter-chip spacing is reconstructed by combining SEM measurements at the overlapping chip edges with profilometer scans of the exposed qubit-chip back side plane. The samples are tilted by $90\si{\degree}$ in the SEM to image the gap between the two silicon substrates. An example is presented in Fig.\,\ref{fig_app2}\,(a). The effective spacing is obtained by subtracting the combined thickness of the two niobium films from the measured silicon-to-silicon distance, $h_\mathrm{SEM}=h_\mathrm{Si-Si}-2t_\mathrm{Nb}$ with $t_\mathrm{Nb}=150~\si{\nano\meter}$. SEM images are acquired at $1~\si{\milli\meter}$ intervals along each chip edge, resulting in nine measurements per edge. The uncertainty of an individual SEM spacing measurement is estimated to be below $100~\si{\nano\meter}$.
 
Profilometer measurements are then used to extract the relative spacing variation of the bonded assembly away from the chip edges. Since the expected deformation is dominated by global rather than local features, measurements are performed at three positions along the chip length ($0$, $4$ and $8~\si{\milli\meter}$). The profilometer measures the height of the exposed top-chip surface relative to an arbitrary reference plane, and the raw profilometer data do not directly provide the absolute chip-to-chip spacing. However, assuming that the $525~\si{\micro\meter}$ thick silicon substrate does not significantly compress or expand during bonding, the top-chip surface provides a representative measure of the deformation occurring at the bonding interface. The profilometer profiles are therefore linearly re-calibrated using the SEM-measured spacings at the corresponding chip edges. This converts the relative profilometer height profile into an absolute spacing profile, as represented in Fig.\,\ref{fig_app2}\,(b).
 
After this re-calibration, the SEM and profilometer data are combined and fitted with a quadratic surface,
\begin{equation}
    z(x,y)=a_0+a_xx+a_yy+a_{xx}x^2+a_{yy}y^2+a_{xy}xy
\end{equation}
where $z(x,y)$ is the local inter-chip spacing and $(x,y)$ are the in-plane chip coordinates, as shown in Fig.\,\ref{fig_app2}\,(c). The use of a quadratic surface assumes that the dominant chip-scale deformation consists of a combination of global tilt and bowing of the silicon chips. The fitted surface provides a two-dimensional reconstruction of the spacing across the bonded region. The mean chip-to-chip spacing and the spatial standard deviations are extracted from this reconstructed map. The mean spacing is calculated as the average of $z(x,y)$ over the bonded area, while the spatial variation is quantified by the standard deviation of $z(x,y)$ over the same area. 
 
To extract the global tilt, the SEM and profilometer data are additionally fitted with a plane,
\begin{equation}
    z_\mathrm{plane}(x,y)=b_0+b_xx+b_yy.
\end{equation}
The polar tilt angle $\theta$ is obtained from the magnitude of the out-of-plane gradient,
\begin{equation}
\theta =
\arctan\left(\sqrt{b_x^2+b_y^2}\right).
\end{equation}
The azimuthal angle $\phi$, defining the in-plane direction of the tilt, is given by
\begin{equation}
    \phi=\arctan2(b_y,b_x).
\end{equation}
 
\section{\label{app3}SU-8 spacers deformation}
 
\setcounter{equation}{4}
 
In Sec.~\ref{sec3}, the samples assembled using a multi-angle bonding procedure exhibit an average inter-chip spacing height of $10.04~\si{\micro\meter}$ despite an average spacer thickness of $10.28~\si{\micro\meter}$. We explain this discrepancy by the possible deformation of the SU-8 resist during compression. Assuming a total spacer area of $0.18~\si{\milli\meter\squared}$, as designed for the specific samples, and a uniformly distributed bonding force of $16~\si{\newton}$, each spacer experiences a compressive stress of approximately $88~\si{\mega\pascal}$. For a simple uniaxial compression, the expected spacer deformation is
\begin{equation}
    \Delta h = \frac{F}{A E} h_0 ,
\end{equation}
where $F$ is the applied bonding force, $A$ is the spacer area carrying the load, $E$ is the Young's modulus of SU-8, and $h_0$ is the fabricated spacer height. Using an approximate Young's modulus of $4~\si{\giga\pascal}$ for SU-8 spacers of initial thickness $h_0 = 10.28~\si{\micro\meter}$, this stress corresponds to a compression of approximately $228~\si{\nano\meter}$, in agreement with the measured reduction in spacing~\cite{Xu2016SU8MechanicalBehavior}.
 
\section{\label{app4}DC characterization}
 
\begin{figure}[b]
\includegraphics{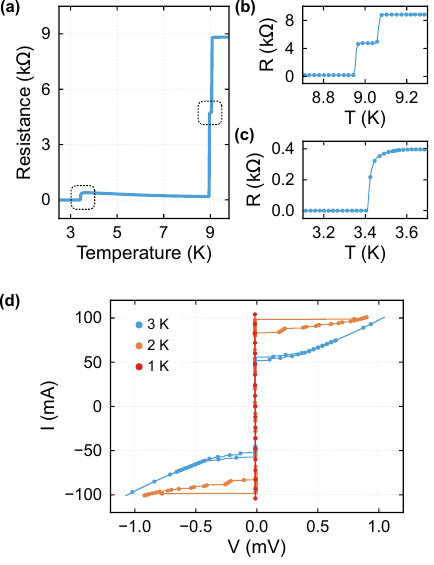}
\caption{Critical temperature and critical current characterization of indium interconnects. (a) Full temperature-resistance curve of an indium daisy-chain. Dotted squares highlight enlarged views. (b) Enlarged view of two niobium transitions near $9~\si{\kelvin}$. (c) Enlarged view of the indium transition at $3.4~\si{\kelvin}$. (d) Current-voltage characteristics measured at different cryogenic temperatures.}
\label{fig_app3}
\end{figure}
 
To assess the electrical performance of the indium interconnects after assembly, we measure the temperature-dependent resistance of an indium daisy-chain comprising 3810 indium interconnects. An SEM picture of a representative indium interconnect is shown in Fig.\,\ref{fig1}\,(e). Resistance measurements are performed in a kiutra adiabatic demagnetization refrigerator, while ramping the temperature from $2~\si{\kelvin}$ to $10~\si{\kelvin}$, as shown in Fig.\,\ref{fig_app3}\,(a). Three superconducting transitions are observed in the resistance curve, highlighted in Fig.\,\ref{fig_app3}\,(b) and (c). Two closely spaced transitions near $9~\si{\kelvin}$, more specifically at $\mathrm{T_{Nb_1}}=9.06~\si{\kelvin}$ and $\mathrm{T_{Nb_2}}=8.95~\si{\kelvin}$, are attributed to slightly different critical temperatures of the niobium films on the control and qubit chips. Since the two films are deposited and processed independently, small variations in their superconducting properties can be expected~\cite{Rairden1964CriticalTemperatureNbTa}. A third transition at $\mathrm{T_{In}}=3.42~\si{\kelvin}$ corresponds to the superconducting transition of indium~\cite{Roberts1976SuperconductiveMaterials}. 
 
Below the indium critical temperature, a total of 500 data points yield an average residual resistance $\mathrm{R_{total}}=(1.43\pm17.77)~\si{\micro\ohm}$, with a standard error of the mean of $0.80~\si{\micro\ohm}$. This corresponds to a resistance $\mathrm{R_{In}}=0.37~\si{\nano\ohm}$ per interconnect, with a standard deviation of $4.67~\si{\nano\ohm}$. Since the mean is far smaller than the scatter, an upper bound is extracted using a $2\sigma$ interval: $R_{\mathrm{In}} < 0.37 + 2\times 4.67 = 9.71~\si{\nano\ohm}$, comparable to existing literature~\cite{Rosenberg2017_3DIntegrated, Kosen2022BuildingBlocks}. This result confirms the fabrication of low-resistance galvanic indium interconnects between the control and qubit ground planes. To further assess their suitability for use in flux-bias lines for flux-tunable elements, the indium interconnects are characterized by measuring current-voltage curves at cryogenic temperatures. Current-voltage characteristics of a daisy-chain made of 10 indium interconnects are measured at $1~\si{\kelvin}$, $2~\si{\kelvin}$, and $3~\si{\kelvin}$, as shown in Fig.\,\ref{fig_app3}\,(d). At $2~\si{\kelvin}$ and $3~\si{\kelvin}$, the return current $I_r$ is smaller than the switching current $I_c$, resulting in hysteretic current-voltage characteristics. This behavior is attributed to local heating in the resistive state and is consistent with previous observations~\cite{Uchihashi2011Macroscopic}. At $1~\si{\kelvin}$, no switching event is resolved within the accessible measurement range; the daisy-chain remains superconducting up to the maximum applied current, setting a lower bound of $I_c>105~\si{\milli\ampere}$ for $T\leqslant 1~\si{\kelvin}$, exceeding previously reported values and typical flux-bias currents used in superconducting quantum circuits~\cite{Akhtar2016GapTunableFluxQubit, Rosenberg2017_3DIntegrated, Lucas2024IndiumBumpBonding}. 
 
\section{\label{app5}Circuit simulations}
 
Eigenmode simulations are performed using Ansys HFSS to determine the resonant frequencies of the $\lambda/4$ CPW readout resonators used in this work. The simulation model consists of a resonator on top of a silicon substrate and inductively coupled to a feedline in either a planar or flip-chip configuration. The silicon substrate is modeled as a dielectric with a relative permittivity of $\epsilon_r = 11.45$, while all metallic structures are treated as perfect electric conductors (PECs). The convergence limit for these simulations is chosen to be below 1\%.
 
The transmon anharmonicity is estimated by combining electrostatic simulations with the transmon energy spectrum. The qubit geometry is simulated in Ansys Maxwell with the qubit pads placed on a silicon substrate. The two capacitor pads are assigned potentials of $+1~\si{\volt}$ and $-1~\si{\volt}$, respectively, while the surrounding ground plane is held at 0 V. From the resulting capacitance matrix, the effective qubit capacitance $C_q$ is obtained, which then gives the charging energy $E_c = e^2/(2C_q)$. Together with the Josephson energy $E_J$, this determines the transmon energy spectrum. We then calculate the exact transition frequencies $f_{01}$ and $f_{12}$ from the Mathieu-function solution of the transmon Hamiltonian and use this to calculate anharmonicity $\alpha = f_{12}-f_{01}$ as expected from simulations~\cite{Koch2007Transmon}.
 
\section{\label{app6}RF measurement setup}
 
\begin{figure}[b]
\includegraphics{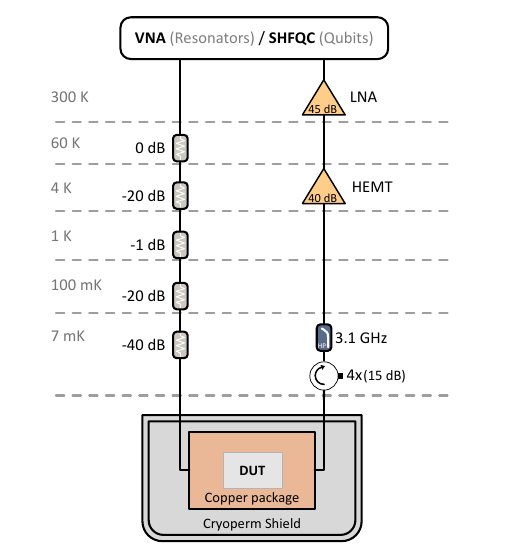}
\caption{RF measurement setup for both resonator and qubit devices. For details, see main text.}
\label{fig_app4}
\end{figure}

The RF-characterization samples are wire-bonded to a gold-coated printed circuit board (PCB), mounted inside a copper package, and installed on the mixing chamber stage of a dilution refrigerator. The package is enclosed by two Cryoperm shields to suppress stray magnetic fields. The electronic wiring from room temperature to the millikelvin stage is shown in Fig.\,\ref{fig_app4}. Resonator measurements are performed in a Bluefors Fast Sample Exchange (FSE) cryostat using a Keysight vector network analyzer (VNA), while qubit control and readout are carried out in a Bluefors LD system with a Zurich Instruments quantum controller (SHFQC). 
 
The input lines provide a total attenuation of \SI{81}{\decibel}. In addition, the insertion loss of the input cabling is characterized independently, yielding an additional attenuation of approximately \SI{8}{\decibel} at \SI{4}{\giga\hertz}. This frequency-dependent loss is taken into account when estimating the photon number in the $\lambda/4$ resonators. In the Bluefors FSE cryostat, the output signal is routed through a cryogenic switch, whereas the Bluefors LD cryostat does not include a switch. Two dual junction isolators provide unidirectional signal routing, while a high-pass filter (Mini-Circuits VHF-3100+) suppresses low-frequency noise. The transmitted signal is amplified by a HEMT amplifier at the \SI{4}{\kelvin} stage, followed by a low-noise amplifier (LNA) at room temperature.

\begin{figure*}[t]
\includegraphics{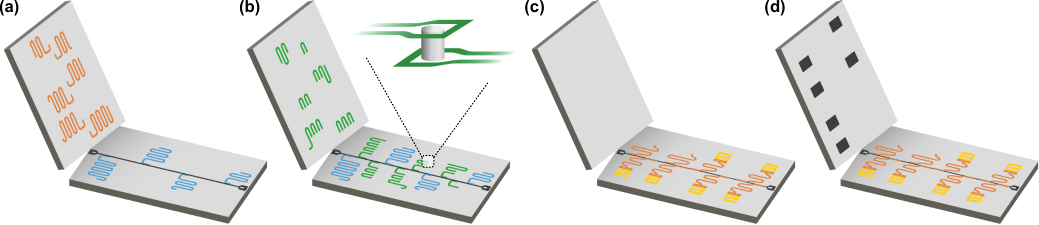}
\caption{3D schematics of the resonator and qubit samples characterized in this article. For the qubit samples, exceptionally, the larger bottom chip follows a qubit-chip fabrication flow while the smaller top chip undergoes a control-chip fabrication flow. This enables planar characterization of the qubits prior to flip-chip assembly. (a) Sample with resonators placed either on the control- or the qubit-chip, colored in blue and orange respectively. (b) Sample composed of resonators with indium transitions as well as four reference resonators on the control chip, shown in green and blue respectively. An additional schematic of an indium transition in a CPW is included. (c) Sample with qubits facing a typical niobium ground plane, the capacitor pads are represented in yellow and the readout resonator in orange. (d) Sample composed of qubits facing an etched silicon pocket, again the capacitor pads are represented in yellow and the readout resonator in orange.}
\label{fig_app5}
\end{figure*}

\section{\label{app7}RF samples}

\setcounter{equation}{5}
 
Fig.\,\ref{fig_app5} (a) and (b) illustrate the resonator geometries used to evaluate microwave losses in the flip-chip architecture. In the first design, shown in Fig.\,\ref{fig_app5}\,(a), the $\lambda/4$ CPW resonators are fabricated on either the control chip or the qubit chip and capacitively coupled to a feedline on the control chip. In the second design, shown in Fig.\,\ref{fig_app5}\,(b), the CPW resonators are distributed between the two chips and capacitively coupled to a feedline on the control chip. At each indium transition, the center conductor is made galvanically continuous through an indium interconnect, allowing the resonator mode to be routed between the control and qubit chips. The number of indium transitions in the resonator center conductor is denoted by $n_c$; for $n_c = 2$ the center conductor transfers from the control chip to the qubit chip and then back to the control chip.

Fig.\,\ref{fig_app5}\,(c) and (d) show the flip-chip qubit geometries fabricated for the parameter-targeting and coherence measurements. To characterize the qubit devices both before and after flip-chip assembly, the qubit structures are designed such that the complete circuit can be measured in a planar configuration prior to bonding. Therefore, the feedlines are fabricated on the same chip as the readout resonators, capacitor pads, Josephson junctions, and indium pads. The opposing chip contains only the structures required for the flip-chip integration, namely the SU-8 spacers and indium bumps. This configuration enables the qubit chips to be characterized in their planar geometry before assembly and subsequently remeasured after flip-chip bonding, providing a direct comparison of the circuit parameters and coherence before and after integration.
 
The main text focuses on the niobium-facing geometry shown in Fig.\,\ref{fig_app5}\,(c), in which the qubit capacitor pads face the niobium ground plane of the control chip. We also fabricate a silicon-facing variant, shown in Fig.\,\ref{fig_app5}\,(d), where windows are etched into the control-chip ground plane so that the qubit capacitor pads face exposed silicon. This geometry has been proposed to reduce metal--air (MA) interface participation, potentially improving coherence and reducing sensitivity to inter-chip spacing variations~\cite{Li2023OptimizingResonatorFrequencyStability, Luo2025MultifloatingFlipChipQubits}. It is included here to demonstrate the compatibility of the fabrication process with different control-chip surface configurations.
 
\begin{table*}[!t]
\centering
 
\caption{Comparison of parameter errors between simulation and measurement for the resonator frequency, qubit frequency, and charging energy of six transmon qubits before and after flip-chip assembly. The qubits are first characterized in a planar geometry and subsequently bonded using the multi-angle flip-chip process before being remeasured. Here, $\epsilon_x=100\times(x_\mathrm{meas}-x_\mathrm{sim})/x_\mathrm{sim}$ denotes the signed percentage deviation from the simulated target value.}
 
\vspace{10pt}
 
(a) Qubits facing niobium \\ [6pt]
 
\renewcommand{\arraystretch}{1.5}
\setlength{\tabcolsep}{5pt}
 

\begin{tabular}{|@{\hspace{3mm}}c@{\hspace{3mm}}|SS@{\hspace{3mm}}|SS@{\hspace{3mm}}|SS@{\hspace{3mm}}|SS@{\hspace{3mm}}|SS@{\hspace{3mm}}|SS|}\hline
 
\multirow{3}{*}{\textbf{Error}}
& \multicolumn{2}{c|}{\textbf{Q1}}
& \multicolumn{2}{c|}{\textbf{Q2}}
& \multicolumn{2}{c|}{\textbf{Q3}}
& \multicolumn{2}{c|}{\textbf{Q4}}
& \multicolumn{2}{c|}{\textbf{Q5}}
& \multicolumn{2}{c|}{\textbf{Q6}}
\\
 
& \multicolumn{1}{c}{\phantom{-}2D} & \multicolumn{1}{c|}{3D}
& \multicolumn{1}{c}{\phantom{-}2D} & \multicolumn{1}{c|}{3D}
& \multicolumn{1}{c}{\phantom{-}2D} & \multicolumn{1}{c|}{3D}
& \multicolumn{1}{c}{\phantom{-}2D} & \multicolumn{1}{c|}{3D}
& \multicolumn{1}{c}{\phantom{-}2D} & \multicolumn{1}{c|}{3D}
& \multicolumn{1}{c}{\phantom{-}2D} & \multicolumn{1}{c|}{3D}
\\
 
\hline
 
$\bm{\epsilon}_{\mathbf{f_r}}$ & -3.64 & -3.40 & -3.89 & -3.84 & -4.44 & -3.63 & -4.74 & -4.00 & -5.08 & -4.24 & -4.96 & -4.41 \\
$\bm{\epsilon}_{\mathbf{f_q}}$ & -2.88 & -1.31 & -0.79 & \phantom{-}0.19 & -4.48 & -3.76 & -2.90 & -2.16 & -1.60 & -1.04 & \phantom{-}0.18 & \phantom{-}0.84 \\
$\bm{\epsilon_{E_c}}$ & -1.54 & -0.10 & -1.77 & -0.69 & -1.26 & -1.40 & -1.55 & -1.46 & -1.34 & -1.75 & -1.62 & -1.70 \\
 
\hline
\end{tabular} \\ [10pt]
 
(b) Qubits facing silicon \\ [6pt]

 
\begin{tabular}{|@{\hspace{3mm}}c@{\hspace{3mm}}|SS@{\hspace{3mm}}|SS@{\hspace{3mm}}|SS@{\hspace{3mm}}|SS@{\hspace{3mm}}|SS@{\hspace{3mm}}|SS|}\hline
 
\multirow{3}{*}{\textbf{Error}}
& \multicolumn{2}{c|}{\textbf{Q1}}
& \multicolumn{2}{c|}{\textbf{Q2}}
& \multicolumn{2}{c|}{\textbf{Q3}}
& \multicolumn{2}{c|}{\textbf{Q4}}
& \multicolumn{2}{c|}{\textbf{Q5}}
& \multicolumn{2}{c|}{\textbf{Q6}}
\\
 
& \multicolumn{1}{c}{\phantom{-}2D} & \multicolumn{1}{c|}{3D}
& \multicolumn{1}{c}{\phantom{-}2D} & \multicolumn{1}{c|}{3D}
& \multicolumn{1}{c}{\phantom{-}2D} & \multicolumn{1}{c|}{3D}
& \multicolumn{1}{c}{\phantom{-}2D} & \multicolumn{1}{c|}{3D}
& \multicolumn{1}{c}{\phantom{-}2D} & \multicolumn{1}{c|}{3D}
& \multicolumn{1}{c}{\phantom{-}2D} & \multicolumn{1}{c|}{3D}
\\
 
\hline
 
$\bm{\epsilon}_{\mathbf{f_r}}$ & -3.69 & -3.41 & -3.92 & -3.66 & -4.56 & -3.94 & -4.77 & -4.03 & -5.13 & -4.20 & -5.51 & -4.24 \\
$\bm{\epsilon}_{\mathbf{f_q}}$ & -3.67 & -0.65 & -1.58 & \phantom{-}1.56 & \phantom{-}2.22 & \phantom{-}4.72 & -3.85 & -1.20 & -2.15 & \phantom{-}0.92 & -0.85 & \phantom{-}2.06 \\
$\bm{\epsilon_{E_c}}$ & -1.34 & -0.21 & -1.53 & \phantom{-}0.35 & -2.17 & -0.62 & -1.32 & -0.25 & -1.48 & -0.17 & -1.81 & -0.41 \\
 
\hline
\end{tabular} 
 
\label{tab_app1}
\end{table*}

\begin{table*}
\centering
 
\caption{$T_1$ and $T_{2,\mathrm{echo}}$ measurements of single transmon qubits for different geometries. The measurements are performed continuously over a 24-hour period.}
 
\vspace{10pt}
 
(a) Reference planar qubits \\ [6pt]
 
\renewcommand{\arraystretch}{1.5}
\setlength{\tabcolsep}{5pt}
 

\begin{tabular}{|@{\hspace{5mm}}c@{\hspace{5mm}}|
                 @{\hspace{5mm}}c@{\hspace{5mm}}|
                 @{\hspace{5mm}}c@{\hspace{5mm}}|
                 @{\hspace{5mm}}c@{\hspace{5mm}}|
                 @{\hspace{5mm}}c@{\hspace{5mm}}|
                  @{\hspace{5mm}}c@{\hspace{5mm}}|
                 @{\hspace{5mm}}c@{\hspace{5mm}}|}\hline
 
\textbf{Parameters}
& \textbf{Q1}
& \textbf{Q2}
& \textbf{Q3}
& \textbf{Q4}
& \textbf{Q5}
& \textbf{Q6}
\\
 
\hline
 
$\bm{T_1}$ [\si{\micro\second}] & $145 \pm 30$ & $189 \pm 37$ & $225 \pm 42$ & $157 \pm 44$ & $132 \pm 16$ & $152 \pm 47$ \\
$\bm{T_{2,\mathrm{echo}}}$ [\si{\micro\second}] & $273 \pm 83$ & $355 \pm 83$  & $390 \pm 96$ & $309 \pm 100$ & $240 \pm 45$ & $267 \pm 83$\\

\hline
\end{tabular} \\ [10pt]
 
(b) Flip-chip qubits facing niobium\\ [6pt]
 
 
\begin{tabular}{|@{\hspace{5mm}}c@{\hspace{5mm}}|
                 @{\hspace{5mm}}c@{\hspace{5mm}}|
                 @{\hspace{5mm}}c@{\hspace{5mm}}|
                 @{\hspace{5mm}}c@{\hspace{5mm}}|
                 @{\hspace{5mm}}c@{\hspace{5mm}}|
                 @{\hspace{5mm}}c@{\hspace{5mm}}|
                 @{\hspace{5mm}}c@{\hspace{5mm}}|}\hline
 
\textbf{Parameters}
& \textbf{Q1}
& \textbf{Q2}
& \textbf{Q3}
& \textbf{Q4}
& \textbf{Q5}
& \textbf{Q6}
\\
 
\hline
 
$\bm{T_1}$ [\si{\micro\second}] & $176 \pm 15$ & $165 \pm 27$ & $137 \pm 22$ & $169 \pm 16$ & $150 \pm 20$ & $156 \pm 20$  \\
$\bm{T_{2,\mathrm{echo}}}$ [\si{\micro\second}] & $335 \pm 36$ & $316 \pm 65$ & $223 \pm 59$ & $320 \pm 43$ & $266 \pm 35$ & $304 \pm 39$ \\
 
\hline
\end{tabular} \\ [10pt] 
 
(c) Flip-chip qubits facing silicon \\ [6pt]
 

\begin{tabular}{|@{\hspace{5mm}}c@{\hspace{5mm}}|
                 @{\hspace{5mm}}c@{\hspace{5mm}}|
                 @{\hspace{5mm}}c@{\hspace{5mm}}|
                 @{\hspace{5mm}}c@{\hspace{5mm}}|
                 @{\hspace{5mm}}c@{\hspace{5mm}}|
                 @{\hspace{5mm}}c@{\hspace{5mm}}|
                 @{\hspace{5mm}}c@{\hspace{5mm}}|}\hline
 
\textbf{Parameters}
& \textbf{Q1}
& \textbf{Q2}
& \textbf{Q3}
& \textbf{Q4}
& \textbf{Q5}
& \textbf{Q6}
\\
 
\hline
 
$\bm{T_1}$ [\si{\micro\second}] & $159 \pm 26$ & $141 \pm 45$ & $173 \pm 23$ & $187 \pm 31$ & $138 \pm 41$ & $169 \pm 34$  \\
$\bm{T_{2,\mathrm{echo}}}$ [\si{\micro\second}] & $154 \pm 38$ & $241 \pm 76$ & $340 \pm 51$ & $298 \pm 43$ & $254 \pm 81$ & $292 \pm 63$ \\
 
\hline
\end{tabular} 
\label{tab_app2}
\end{table*}
 
The individual circuit-parameter deviations are listed in Table~\ref{tab_app1}. The table reports the individual deviations for the readout-resonator frequency $f_r$, qubit frequency $f_q$, and charging energy $E_c$. For the niobium-facing device, these values underlie the averaged parameter-targeting comparison discussed in the main text. The silicon-facing data are provided for completeness and show parameter deviations of comparable magnitude, remaining at the few-percent level before and after flip-chip assembly. This suggests that, to first order, the multi-angle bonding procedure sufficiently mitigates the effects of residual tilt, rendering the possible benefit of an etched silicon window negligible in this regard.
 
The individual relaxation and echo-dephasing times are listed in Table~\ref{tab_app2}. For each sample, $T_1$ and $T_{2,\mathrm{echo}}$ are repeatedly measured over a 24-hour period. The reported values correspond to the mean and standard deviation of the repeated measurements for each individual qubit. Again, no clear trend is noticeable between the two different qubit geometries. This indicates that within the range of coherence times measured here, differences in surface participation ratios do not produce a clearly discernible effect on coherence.

\clearpage

\newpage

\bibliography{bib}

\end{document}